# Modeling High Quantum Bit Rate QKD Systems over Optical Fiber

Michal Mlejnek*[a], Nikolay A. Kaliteevskiy[b], Daniel A. Nolan[a]
[a]Corning Incorporated, Corning, NY 14831; [b]Corning Scientific Center, Saint Petersburg, Russia

## ABSTRACT

There is considerable interest in finding conditions under which the quantum key distribution (QKD) propagation distances over fiber and secure key rate (SKR) are maximized for a given acceptable quantum bit error rate. One way to increase the secure key rate is to increase quantum bit rate, i.e. use shorter pulses. Short pulses propagating in a fiber are subject to temporal broadening caused by chromatic dispersion (CD) which leads to inter-symbol-interference and quantum bit-error rate increase. Current commercial QKD systems employ 1 Gb/s quantum bit rate sources, and the transition to 10 Gb/s system is being researched. While not very important in the 1 Gb/s, the effect of CD cannot be neglected in 10 Gb/s or higher quantum bit rate systems.

## 1. INTRODUCTION

Modern long-haul terrestrial and submarine communication systems deployed now use coherent transmission at 100 Gb/s or higher. Chromatic dispersion in classical channels is compensated digitally (i.e., no optical dispersion compensation within the optical link). While most current field studies of simultaneous quantum and classical signal transmission were performed in (often) higher loss dark fiber [1]-[9], low loss fiber shows clear benefit in classical transmission as well as in practical installation of QKD systems over fiber [10]. Besides higher loss, the dark fiber also often does not provide the low optical nonlinearity medium required for efficient, high capacity modern "classical" coherent systems – to that end, large effective area fibers, such as quasi-single-mode fiber [11], are intensively studied. Record setting transmission distances over 300 km [12] and reaching 400 km [13] (not necessarily at the high secure key rates) were achieved using low loss fiber. Hence, low loss and low nonlinearity fibers are of interest when high bit rate classical and quantum channel transmission over the same fiber is of interest [14].

Some of the work cited above is permeated with the effort to find conditions under which the fiber can be used for both QKD and classical signal transmission simultaneously. The main challenge is the contamination of a very weak QKD signal, which typically uses an average number $\mu$ of photons per pulse smaller than 1, by a strong classical optical signal pulse containing approximately 100 photons per pulse on average or more for a Gb/s or higher bit rate classical link without amplification. While linear crosstalk (LCXT) can be battled by introducing spectral filtering, fiber optical nonlinearities cause inelastic scattering of photons that generates photons in the spectral region used by the QKD signal. The most common nonlinear impairments are spontaneous Raman scattering (SpRS) and four-way mixing (FWM) nonlinearity, which have been studied previously, e.g. [1],[6],[15]. We previously reviewed these impairments on the quantum channel performance in the framework of individual attacks [16], including the advantageous impact of forward error correction (FEC) and coherent modulation formats (CMF) on the transmission of the classical signal.

There is an additional impairment that becomes relevant for quantum bit rates on the order larger than 10 Gb/s for signals propagating through dispersive media, chromatic dispersion (CD). As mentioned above, QKD signal transmission uses very weak signals containing less than one photon per pulse – quantum nature of signal precludes the use of electronic CD compensation; only analog optical CD compensation within the optical link is possible. The basics of the effect of CD for bit rates have been laid down in Ref. [16].

In this paper, we shall highlight few features of QKD fiber transmission system performance that are associated with CD, including a look at the effect of CD pre-compensation. CD pre-compensation has been chosen since it does not introduce additional loss to the transmission of the signal. As opposed to the CD post-compensation that happens after the fiber link and the CD compensation that happens within the fiber link, the pre-compensation takes place before the QKD signal is attenuated to the required sub-photon count and sent through the transmission fiber.

Basic idea of CD pre-compensation relies on input quantum pulse chirp engineering, which results in the shortest pulse arriving at the quantum detector at a specified transmission fiber length. To (optically) pre-compensate CD, one can apply dispersion compensation fibers (DCF). The typical parameters of DCF are 0.42 dB/km attenuation and -132.4 ps/nm/km CD [17] (see Table 2). For example, the length of DCF that compensates 300 km of Corning® Vascade® EX2000 optical fiber with $D_{EX2000}$ = 20.35 ps/nm/km dispersion is calculated as

$$L_{DCF} = \frac{L_{EX2000} \cdot D_{EX2000}}{|D_{DCF}|} \approx 46 \text{ km}, \tag{1.1}$$

which leads to ~19 dB DCF attenuation. Since large attenuation may be a problem for signal power control, alternative means of CD pre-compensation utilizing fiber Bragg grating (FBG) or chirped Bragg grating [18] can be used. The advantage of these methods is a much smaller attenuation [18],[19] – an insertion loss < 0.5 dB is reported in Ref. [19] – which may make them also useful for CD post- and in-line compensation.

## 2. BASIC SETUP AND PARAMETERS

Our goal is to understand the dependence of the performance of high bit rate classical/quantum wavelength-division-multiplexing (WDM) transmission on the fiber and classical channel modulation format parameters. QKD channel performance is characterized by quantum bit-error rate (QBER) and secure key rate. We use the basic setup of Ref. [6], in which a Cerberis QKD system (from ID Quantique [20]) assumes four classical communication channels set up in addition to the quantum communication channel. Two classical channels are required for the bidirectional encrypted data transmission between the encryptors and an additional two classical authenticated channels are required for distillation, i.e. key sifting, error correction and privacy amplification, one from Alice to Bob and one from Bob to Alice.

### 2.1 Quantum channel characteristics

The quantum transmission part is characterized by the frequency $f_{rep}$ with which Bob generates the pulses. The pulses are sent to Alice, who uses them for clock synchronization, attenuates them, modulates them, and sends them back as a quantum signal to Bob. Superconducting nanowire single-photon detectors (SNSPD) are used to detect the quantum signal. Important SNSPD characteristics are the gate time (i.e., temporal filtering) $\Delta t_{gate}$, dead time $\tau_{dead}$, after-pulse probability to the total detection probability fraction $\rho_{AP}$, and quantum detection efficiency $\eta$. Before being detected, the quantum signal experiences an internal loss $t_B$ due to optical components in Bob's detection system. The interference visibility $V$ is measured in order to determine QBER. As in Ref. [16], we use SKR formulas for individual attacks for COW QKD protocols [20],[21] in the secret key distillation phase, despite the existence of analysis of general or coherent attack formulas for decoy-state BB84 [22],[23] (currently, the most popular QKD protocol). We assume that quantum error correction performance is similar to the CASCADE algorithm [6]. After privacy amplification Alice and Bob remain with shared secret keys.

We assume that a permanent advanced encryption standard AES-256 (key length = 256 bits, working with 128 bits long blocks [24]), is used to encrypt transmission data links between Alice and Bob via a pair of Ethernet encryptors. Since the practical period between secret key updates is currently about one minute [25] and standard coherent transponders have 25 Gbaud (with no FEC overhead, commercially 32 Gbaud with FEC overhead) information symbol rate that corresponds to 50 and 100 Gb/s capacity for PM-BPSK or PM-QPSK modulation formats, about 300 and 600 Gb of data is encrypted over classical channel using the same key, respectively. One minute AES-256 refresh rate results in SKR of approximately 8.6 b/s (i.e., 4.3 b/s per encrypted classical channel). Typically, when SKR drops below that limit, the key refresh rate is temporarily reduced to assure continuous operation at the expense of lowering the security level. In order to keep the security level over one channel the same as the bit rate in the classical signal increases, either the refresh rate or SKR threshold for one quantum channel would have to increase proportionally [16].

Similar to the theoretical treatment in Ref. [6], our secret key rate estimation is optimistic, since we ignore any interruptions of the key exchange during key distillation and fiber length measurements. This approximation is more severe at higher key rates that typically occur for short fiber lengths.

### 2.2 Classical channel characteristics

In our analysis we consider four or more classical communication channels that are implemented using standard optical 100 Gb/s WDM transceivers with varying CMF, such as PM-QPSK, DPSK, or 16QAM [26]. The corresponding receiver sensitivity is denoted by $R_x$ and it characterizes the minimum optical signal power incident on the detector such that the signal can be corrected and recovered with less than a given small final bit error ratio (BER); a common choice is $10^{-12}$ [24],[26].

Standard single-mode fiber of different span lengths is used as a fiber link of length $L$. The average fiber attenuation $\alpha$ is characterized in dB/km. For practical reasons, all classical channels are chosen to have larger wavelength than quantum to take advantage of a lower SpRS noise on the anti-Stokes side of the Raman spectrum. SpRS detection probabilities, $p_{ram,b}$, $p_{ram,f}$, and the total SpRS detection probability $p_{ram}$ can be calculated using

$$p_{ram,f(b)} = \frac{P_{ram,f(b)}}{E_{photon}} \cdot \eta \cdot \Delta t_{gate}, \quad p_{ram} = p_{ram,f} + p_{ram,b}, \text{ where } E_{photon} = \frac{hc}{\lambda},$$

$$P_{ram,b} = N_b \cdot P_{out} \cdot \frac{\sinh(\alpha L)}{\alpha} \cdot \rho(\lambda) \cdot \Delta\lambda, \quad P_{ram,f} = N_f \cdot P_{out} \cdot L \cdot \rho(\lambda) \cdot \Delta\lambda, \quad P_{out} = P_{in} \cdot e^{-\alpha L},$$

(2.1)

and index $b$ stands for "backward, $f$ for "forward", $N_b$ and $N_f$ denote the number of backward and forward classical channels, $\Delta\lambda$ denotes quantum receiver bandwidth, and the optical power at the fiber input is denotes as $P_{in}$.

Furthermore, any practical WDM system suffers from insertion loss (IL) and LCXT. The total IL $t_{IL}$ takes into account optical filtering in the WDM, as well as any spectral filtering introduced by additional components or misalignments in the system. LCXT depends on the isolation, $t_{i,a}$, of adjacent WDM channels provided by the WDM filters. To lower the LCXT effect on the quantum channel while using the same WDM filter, we separate the quantum channel by at least twice the classical channel spacing from the next nearest classical channel and denote its LCXT by $t_{i,n-a}$.

The focus of this study is how CD affects high quantum bit rate QKD. We shall lay out the basic principles used to evaluate the effect of temporal pulse spreading due to CD on QKD performance in more detail in Section 3.

## 2.3 Model parameters

Table 1 summarizes all the parameters and the range of their numerical values used in the current study.

Table 1. Model parameters

| Variable | Symbol | Units | Typical value used |
|---|---|---|---|
| fiber attenuation | $\alpha$[dB/km] | dB/km | 0.16, 0.185, 0.195, 0.21, 0.42 |
| fiber length | $L$ | km | 1 – 400 |
| fiber transmission | $t_F$ | - | exp(-$\alpha L$) |
| fiber dispersion | $D$ | ps/nm/km | 0.1, 4.25, 17, 20.35, -132.4 |
| reduction of detection rate imposed by the detection protocol synchronization requirements | $\eta_{duty}$ | - | 0.71 [16] |
| loss of receiver internal components | $t_B$[dB] | dB | 2.65 |
| WDM insertion loss (including optical filtering in the receiver) – "increases the cross-talk by the same amount" | $t_{IL}$[dB] | dB | 1.95 |
| isolation of non-adjacent channels | $t_{i,n-a}$[dB] | dB | 82 |
| isolation of adjacent channels | $t_{i,a}$[dB] | dB | 59 |
| effective Raman cross-section | $\rho(\lambda)$ | (km.nm)$^{-1}$ | $2\times10^{-9}$ |
| quantum receiver bandwidth | $\Delta\lambda$ | nm | 0.6 |
| assumed channel spacing | $\Delta\lambda_{ch}$ | nm | 0.8 |
| (classical) signal bit rate (permanent AES-256 transmission rate; data in one stream) | $f_{AB}$ | Gb/s | 1 – 200 |
| maximum secret key refresh period (fixed) | $T_{AES,max}$ | s | 60 |
| minimum secret key rate required for $N$ channel AES-256 encryption updated once in 60 s (fixed refresh rate) | $f_{AB,min}$ | b/s | $N\times256/60 \sim 4.27N$ |
| classical signal receiver sensitivity (depends on $f_{AB}$) | $R_x$ | dBm | -50 |
| numbers of forward and backward classical channels | $N_f$ $N_b$ | - - | 2 (1 distillation, 1 data encryption) 2 (1 distillation, 1 data encryption) |
| classical channel power output from the fiber | $P_{out}$ | dBm | $R_x$[dBm] + $t_{IL}$[dB] (Ref. [6],[16]) |
| photon energy | $E_{photon}$ | J | $1.278818\times10^{-19}$ |
| CASCADE error correction QBER distillation limit | $QBER_{thr}$ | - | 0.09 |
| CASCADE algorithm correction to discarded bits | $\eta_{ec}$ | - | 6/5 |

| number of quantum detectors | $N_d$ | - | 2 |
|---|---|---|---|
| Bob's "quantum" pulse generation rate | $f_{rep}$ | GHz | less or equal to 10 (optimized) |
| Bob's "quantum" pulse generation period | $T = 1/f_{rep}$ | ps | 100 or longer (optimized) |
| dead time of quantum detector | $\tau_{dead}$ | μs | 0.1 for SNSPD |
| quantum detector gate duration time | $\Delta t_{gate}$ | ns | $1/(2f_{rep})$ (optimized) |
| quantum detection efficiency | $\eta$ | - | SNSPD: 0.014 (at $p'_{dc} \sim 50$ s$^{-1}$) |
| factor depending on QKD protocol | $\beta$ | - | $\beta_{COW} = 1$ |
| average number of photons per quantum pulse | $\mu$ | - | numerically optimized |
| signal detection probability | $p_\mu$ | - | Eq. (4.2) |
| dark count probability rate | $p'_{dc}$ | ns$^{-1}$ | $50 \times 10^{-9}$/SNSPD |
| dark count probability | $p_{dc}$ | - | $p'_{dc} \cdot \Delta t_{gate}$ |
| Raman noise detection probability | $p_{ram}$ | - | Eqs. (2.1) |
| crosstalk photon detection probability rate | $p'_{LCXT}$ | ns$^{-1}$ | Ref. [6],[16] |
| crosstalk photon detection probability | $p_{LCXT}$ | - | $p'_{LCXT} \cdot \Delta t_{gate}$ |
| ISI detection probability due to chromatic dispersion | $p_{ISI}$ | - | Eq. (4.1) |
| loss of photons due to part of the pulse falling outside of quantum detector gate duration time | $t_{ISI}$ | - | Eq. (3.3) [$\sim \mathcal{O}(1)$] |
| pulse overlap with neighboring time gates (ISI) | $f_{err}^{(ISI)}$ | - | 0.001 |
| coefficient capturing the reduction of detection rate due to quantum detector dead time $\tau_{dead}$ | $\eta_{dead}$ | - | Ref. [6],[16] |
| after-pulse probability to total detection probability ratio | $\rho_{AP}$ | - | SNSPD: 0 |
| after-pulse detection probability | $p_{AP}$ | - | Ref. [6],[16] |
| fringe visibility | $V$ | - | 0.997 |
| input (into DCF) quantum signal pulse duration | $\tau_{FWHM,0}$ | ps | $0.15T$ |
| Input (into DCF) quantum signal pulse chirp | $C_0$ | - | 0 |

## 3. CHROMATIC DISPERSION

Typical fiber CD starts to affect the performance of ~100 km QKD links for quantum bit rates 10 Gb/s and higher. The optical pulses broaden in time so they can interfere with the neighboring bit. We shall discuss two ways in which (quantum) bit error rate (QBER) at the detector is increased by pulse broadening: (1) the probability of a photon detected in the given bit diminishes (effectively reducing photon average number), and (2) the probability of a photon detected in the neighboring bits increases [so-called intersymbol interference (ISI)].

### 3.1 Pulse shape description and parameters

To estimate the effect of fiber CD in long QKD link on its performance, we assume an input quantum pulse with bandwidth limited Gaussian-shaped input intensity. The pulse's envelope retains its Gaussian shape when propagating a distance $L$ through a linear medium; we write its intensity at point $z$ along the fiber (with special points at the fiber input, $z = 0$, and fiber output, $z = L$) and time $t$ as

$$I(z,t) = \frac{1}{\sqrt{\pi}} \frac{2\sqrt{\ln 2}}{\tau_{FWHM,z}} \exp\left[-4\ln 2 \left(\frac{t}{\tau_{FWHM,z}}\right)^2\right]; \quad z = 0, L. \quad (3.1)$$

Here $\tau_{FWHM,z}$ denotes pulse intensity's full-width-half-max (FWHM) duration at distance $z$. The ratio of FWHM at the end of the fiber and FWHM at the input to the fiber for a Gaussian input pulse with a linear chirp $C$ can be given in terms of fiber dispersion parameters $\beta_2$ or $D$ as [27],[16]

$$\frac{\tau_{FWHM,L}}{\tau_{FWHM,0}} = \sqrt{\left(1 + \frac{C\beta_2 L}{\tau_0^2}\right)^2 + \left(\frac{L}{L_D}\right)^2}; \quad L_D \equiv \frac{\tau_{FWHM,0}^2}{4\ln 2 |\beta_2|} = \frac{\pi}{2\ln 2} \frac{c \cdot \tau_{FWHM,0}^2}{\lambda^2 |D|}, \beta_2 = \frac{\lambda^3}{2\pi c^2} \frac{d^2 n}{d\lambda^2}, D = \frac{2\pi c}{\lambda^2} |\beta_2|. \quad (3.2)$$

The dispersion parameter $\beta_2$ is conventionally given in ps$^2$/km, and is related to the commonly used fiber dispersion parameter $D$ given in ps/nm/km. $L_D$ designates the characteristic dispersion length. In the definition of $\beta_2$, $c$ denotes the speed of light in vacuum in nm/ps and $n$ the effective index of refraction of the propagating fiber mode. We shall assume that the pulse after DCF is Gaussian and enters the transmission fiber with non-zero chirp $C$, opposite in sign to $\beta_2$ of the transmission fiber.

## 3.2 Photon average number reduction and ISI

Figure 1 illustrates the lowering of the average number of photons in the pulse being detected within its gate time slot. The average number of photons per quantum channel pulse arriving within the detector time gate will be diminished by the (multiplicative) amount

$$t_{ISI} \equiv \int_{-\Delta t_{gate}/2}^{\Delta t_{gate}/2} dt\, I(L,t) = \frac{1}{2}\left[erfc\left(-\frac{\sqrt{\ln 2}\Delta t_{gate}}{\tau_{FWHM,L}}\right) - erfc\left(\frac{\sqrt{\ln 2}\Delta t_{gate}}{\tau_{FWHM,L}}\right)\right], \quad erfc(x) \equiv \frac{2}{\sqrt{\pi}}\int_x^\infty dt\, e^{-t^2}. \quad (3.3)$$

ISI accounts for the effect of photons in the pulse falling into the gate time slot allocated to the neighboring bit and causing errors. Using Fig. 1 again, we can argue that the error probability due to ISI imminent from one of the neighboring time bits (only nearest neighbor bit slots are considered) will be proportional to a fraction of the elongated pulse that overlaps with the gate, given by

$$f_{err}^{(ISI)} = \int_{T-\Delta t_{gate}/2}^{T+\Delta t_{gate}/2} dt\, I(L,t) = \frac{1}{2}\left[erfc\left(\frac{\sqrt{\ln 2}}{\tau_{FWHM,L}}(2T - \Delta t_{gate})\right) - erfc\left(\frac{\sqrt{\ln 2}}{\tau_{FWHM,L}}(2T + \Delta t_{gate})\right)\right], \quad (3.4)$$

For current typical QKD link parameters we can write [16]

$$f_{err}^{(ISI)} \approx \frac{1}{2}erfc\left[\frac{\sqrt{\ln 2}}{\tau_{FWHM,L}}\left(\frac{2}{f_{rep}} - \Delta t_{gate}\right)\right], \quad f_{rep} \equiv \frac{1}{T}, \quad (3.5)$$

Now, for example, we may find the maximal achievable quantum bit rate $f_{max} \equiv \max\{f_{rep}\}$ by solving Eq. (3.5) with the selected values $f_{err}^{(ISI)} = 0.001$, $\tau_{FWHM,0} = 0.15T = 0.15/f_{max}$ and $\Delta t_{gate} = T/2 = 0.5/f_{max}$ [16] [assuming that the QBER threshold $QBER_{thr} = 0.09$ [6] and $t_{ISI}$ is of order of 1, $\sim \mathcal{O}(1)$]. From the above considerations, it is clear that the relation between the initial pulse duration $\tau_{FWHM,0}$ and the detection time window $\Delta t_{gate}$ can be treated as an additional optimization parameter. In this study, we do not perform this optimization, keeping the ratio at a constant value.

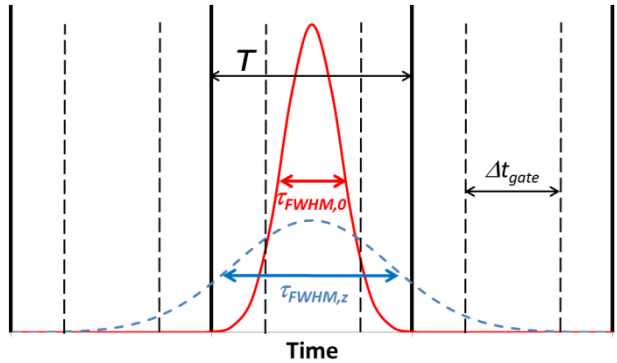

**Figure 1.** Input (full red line) and output (dashed blue line) Gaussian pulses. $T$ – quantum bit period, $\tau_{FWHM,0}$ – width of input pulse, $\tau_{FWHM,L}$ – width of broadened output pulse, $\Delta t_{gate}$ – detection time window.

## 3.3 Fiber parameters

The dispersion parameters of Vascade EX2000 fiber, Corning® LEAF® optical fiber, low dispersion fiber (LDF), also known as dispersion shifted fiber (DSF), standard single mode fiber (we refer to the family of standard single-mode fibers as SMF28e® optical fiber), and DCF fiber are listed in Table 2. Note that parameters used in simulation are rough characteristic of the fiber family, not the specifications associated with the representative fibers.

In Ref. [16], it was presented that the impact of CD in 10 Gb/s QKD systems needs to be accounted for carefully, if fibers with typical *D* are employed. Assuming no CD pre-compensation, fibers with $|D| \leq 0.1$ ps/nm/km were shown to be sufficient to support 10 Gb/s quantum bit rate QKD without significant influence of CD over substantial distances.

Table 2: Fiber properties. (Parameters used in simulation are rough characteristic of the fiber family, not the specifications associated with the representative fibers. In the text, we refer to the fiber families by their representative fibers in this table, even though the modeled parameters do not necessarily agree with their specification for each representative fiber.)

| Fiber family label | Representative fiber | Dispersion @1550nm [ps/nm/km] | Modeled loss [dB/km] |
|---|---|---|---|
| <1> | Corning® Vascade® EX2000 optical fiber | 20.35 | 0.16 |
| <2> | Corning® LEAF® optical fiber | 4.25 | 0.185 |
| <3> | LDF/DSF | 0.1 | 0.185 |
| <4> | SMF28e® optical fiber | 17 | 0.21 |
| <5> | DCF | -132.4 | 0.42 |

## 4. QBER AND SKR IN PRESENCE OF FIBER CD

There are two important characteristics related to the detection of the quantum signal: the rate at which the secret key is delivered to the recipient and the number of bit errors in the sifted key called the QBER, which determines the quality of the quantum signal. We shall use the notation and approximation employed in Refs.[6],[16]

### 4.1 QBER

We introduce the ISI error detection probability due to chromatic dispersion as

$$p_{ISI} = 2 \cdot f_{err}^{(ISI)} \cdot \mu \cdot t_F \cdot t_{IL} \cdot t_B \cdot \eta, \qquad (4.1)$$

where we accounted for contributions from two neighboring time bits and the losses the quantum signal experiences after being coupled to the fiber. The quantum signal detection probability within the given time gate is given by

$$p_\mu = \mu \cdot t_F \cdot t_{IL} \cdot t_B \cdot t_{ISI} \cdot \eta. \qquad (4.2)$$

QBER, the number of errors present in the key obtained after the sifting, can be written, assuming that the visibilities in all the interferometric bases are the same, for COW protocols as [16]

$$QBER_{COW} = \frac{1}{2} \frac{N_d p_{dc} + p_{AP} + p_{ram} + p_{LCXT} + p_{ISI}}{\beta p_\mu + N_d p_{dc} + p_{AP} + p_{ram} + p_{LCXT} + p_{ISI}}. \qquad (4.3)$$

The quantities $p_x$ signify detection probabilities per quantum detector gate. In particular, $p_\mu$ denotes signal, $p_{dc}$ dark count, $p_{AP}$ after-pulse, $p_{ram}$ Raman photon, $p_{LCXT}$ crosstalk photon detection probabilities, and $p_{ISI}$ the ISI error detection probability due to chromatic dispersion. $N_d$ is the number of APD quantum detectors – in our simulations always equals to 2 for COW protocol (equals to the dimension of the bases used in the QKD protocol) [6],[16].

### 4.2 SKR

The SKR calculation starts with the evaluation of the raw detection rate $R_{raw}$ delivered by the detectors due to quantum signals, quantum detector dark counts, after-pulses and additional noise

$$R_{raw} = \left( p_\mu + N_d p_{dc} + p_{AP} + p_{ram} + p_{LCXT} + p_{ISI} \right) f_{rep} \eta_{duty} \eta_{dead}. \qquad (4.4)$$

The coefficient $\eta_{dead}$ is used to account for the reduced detection rate due to a quantum detector dead time $\tau_{dead}$ after each detection, and the coefficient $\eta_{duty}$ is used to characterize the effect of different possible synchronization schemes on the detection rate that is of order of 1, $\sim \mathcal{O}(1)$, but smaller or equal to 1 [6],[16]. A certain fraction of $R_{raw}$ is discarded in the sifting procedure. The sifting algorithm depends on the QKD protocol, as depicted by the parameter $\beta$ that varies from QKD protocol to QKD protocol. The "sifted" key rate is obtained as

$$R_{sift} = \frac{1}{2}\left(\beta p_\mu + N_d p_{dc} + p_{AP} + p_{ram} + p_{LCXT} + p_{ISI}\right) f_{rep} \eta_{duty} \eta_{dead}. \tag{4.5}$$

Denoting $I_{AB}$ and $I_{AE}$ as the mutual information per bit between Alice and Bob, and between Alice and a potential eavesdropper, respectively, the secret key rate $R_{sec}$ after error correction and privacy amplification is estimated as (for incoherent attacks)

$$R_{sec} = R_{sift}(I_{AB} - I_{AE}), \quad I_{AB} = 1 - \eta_{ec} H(QBER), \quad H(p) \equiv -p\log_2 p - (1-p)\log_2(1-p). \tag{4.6}$$

Here, $H(p)$ is the Shannon entropy for a given QBER $p$ that is related to the minimum fraction of bits lost due to error correction. CASCADE error correction algorithm penalty is captured by $\eta_{ec} = 6/5$, a correction to the number of discarded bits, since CASCADE cannot reach the theoretical Shannon limit and since a certain fraction of distilled secret bits is consumed for authentication. $I_{AE}$ depends on the particulars of the algorithm that is used to combat Eve's attacks, e.g., for COW protocols one can estimate [6],[21],[28]

$$I_{AE,COW} = \mu(1-t_F) + (1-V)\frac{1+e^{-\mu t_F}}{2e^{-\mu t_F}}. \tag{4.7}$$

In this expression, the first term corresponds to individual beam splitting attacks and the second to intercept-resend attacks [28], when photon number splitting attacks do not introduce errors. This estimate, which represents the COW protocol security against a large class of collective attacks, assumes that Bob receives at most one photon per bit. Eqs. (4.6) and (4.7) are used to find optimal $\mu$ that maximizes SKR.

## 5. RESULTS

We present two cases of the pre-dispersion use – QKD link with and without classical channels present. First, we study the performance of a nominal 10 Gb/s QKD only system, taking the experimental system described in Ref. [2] and adding CD pre-dispersion. The second study will extend the first one by assuming additional classical channels being transmitted over the same fiber.

### 5.1 "10 Gb/s" QKD system only

Top-left pane in Fig.2 illustrates the limitation of maximal QKD link distance by fiber loss and fiber dispersion [16]. Comparison of LEAF and LDF fibers having the same loss shows that smaller dispersion leads to a larger maximal quantum key rate. LDF/DSF, the fiber with low dispersion, performs the best with only a small margin over Vascade EX2000 fiber and LEAF fiber. Bottom-left pane in Fig.2 shows an idealized QKD performance over fibers with hypothetical zero fiber CD, highlighting the benefit of low loss fibers.

Adding CD pre-compensation enables an increase of QKD link reach, as exhibited at the top-right pane of Fig. 2 for 40 km of DCF and fiber family of Vascade EX2000 fiber. To illuminate the reason, we plot in bottom-right pane of Fig 2 the pulse duration as a function of propagation distance (assuming 15 ps input pulse into DCF and roughly 50 ps detector gate time, and 40 km DCF). We observe QKD link performance increase when the pulse duration reaches detector gate time, as suspected.

Figure 3 demonstrates that the performance of dispersionless fiber at a desired propagation distance can be very nearly recovered by using a specific length of DCF. 50 km DCF pre-dispersion allows the same QKD reach for fiber family of Vascade EX2000 fiber as a dispersionless fiber with the same loss (compare to bottom-left pane in Fig.2).

All simulations were performed by optimizing average photon number $\mu$ and evaluating the ideal $f_{rep}$, with maximal quantum bit rate not larger than 10 Gb/s by solving Eq. (3.5); the input pulse width will also vary as $f_{rep}$ is optimized, because we keep $\tau_{FWHM,0} = 0.15T$. Note that we are not capturing any higher order CD effects beyond those characterized by the dispersion parameter $D$. They are assumed negligible within the single channel spectral width.

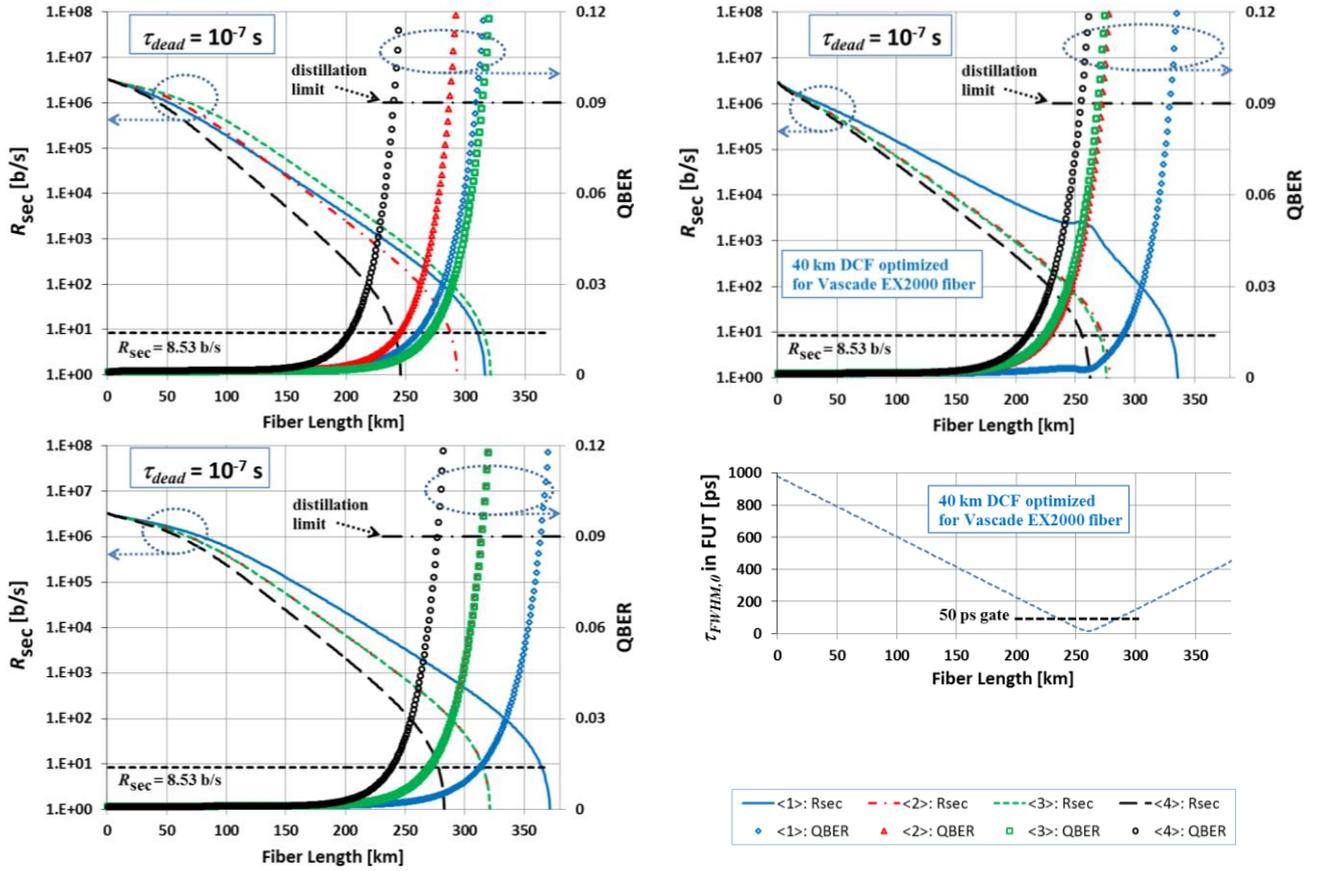

**Figure 2.** Left column: (top) SKR vs. QKD link (only) distance in the case of no CD pre-compensation (0 km DCF); (bottom) all fibers were assumed to have zero fiber CD (no CD pre-compensation needed). Right column: (top) 40 km DCF (CD = -132.4 ps/nm/km) is optimized for pre-compensation of 250 km Vascade EX2000 fiber's fiber family; (bottom) corresponding pulse duration (assuming 15 ps input pulse into DCF). SNSPD were assumed. Ideal $f_{rep}$ and $\mu$ optimized, with maximal quantum bit rate (not larger than 10 Gb/s) allowed by dispersion at each fiber length. Note that $f_{rep}$ is significantly reduced even for short distances to keep QBER below 0.09, but still higher than required by $\tau_{dead}$. *Model parameters are listed in Table 1. Fiber labels are listed in Table 2.*

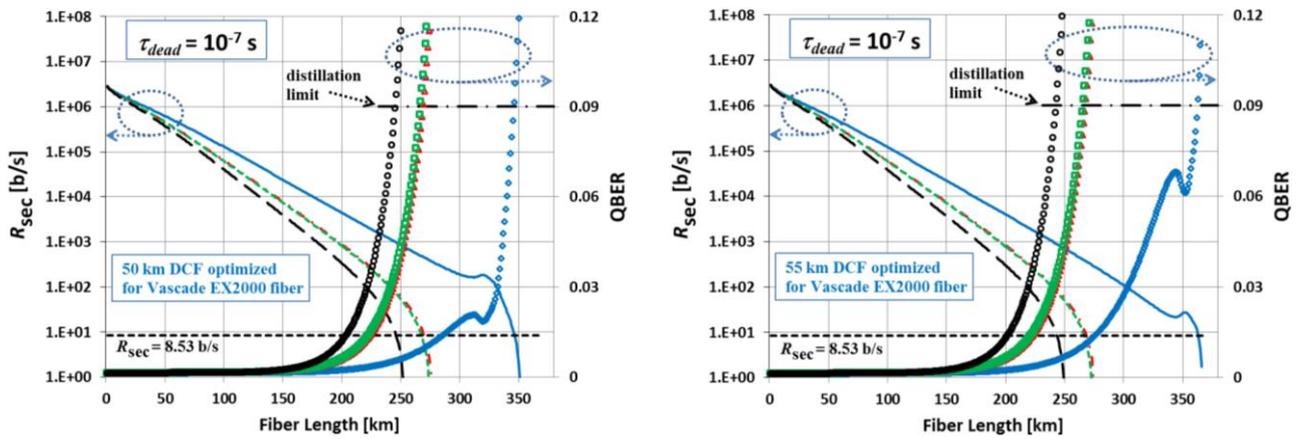

**Figure 3.** Pre-compensation in QKD link only for different lengths of DCF (CD = -132.4 ps/nm/km). Fiber family of Vascade EX2000 fiber was targeted. *Model parameters are listed in Table 1. Legend – see Fig. 2.*

## 5.2 "10 Gb/s" QKD system in presence of classical communication channels

In the second study, we include 4 classical channels, 2 forward (e.g., Bob to Alice) and 2 backward (Alice to Bob), over the same fiber, removed at least 1.6 nm away from the QKD channel. We assume 100 G PM-BPSK modulation format ($R_x$ = -50 dBm) [16]. This contributes LCXT (via $t_{i,n-a}$) and SpRS noise to the quantum signal as discussed in Section 2.2.

Due to these additional impairments, the QKD link reach is significantly shorter than the reach of QKD link only, as can be seen by comparing top-left pane of Fig. 2 with left pane of Fig.4. The length of DCF has to be designed to compensate CD of this reduced reach. It turns out that 30 km DCF pre-dispersion allows nearly the same QKD reach for fiber family of Vascade EX2000 fiber as a dispersionless fiber with the same loss – see the right pane of Fig. 4.

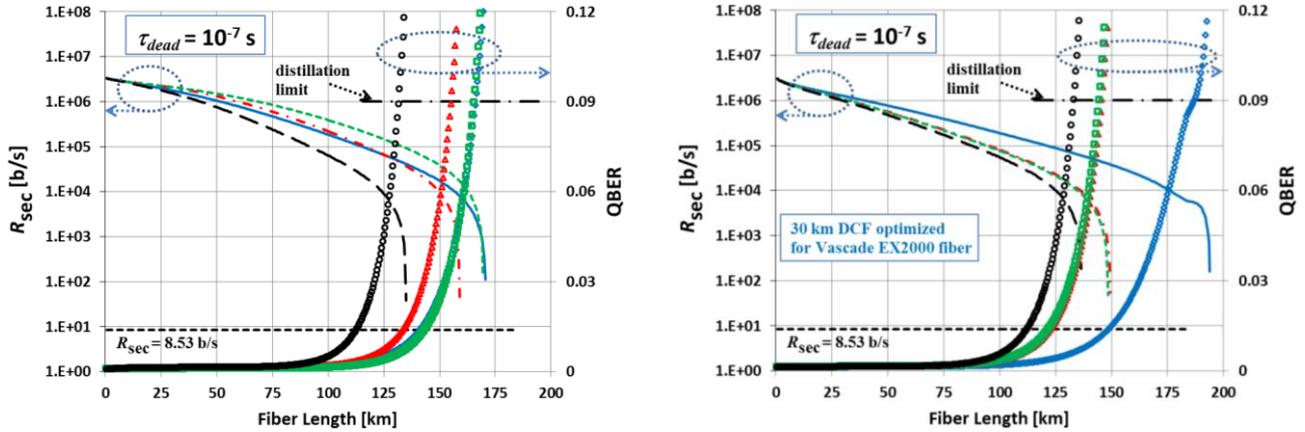

**Figure 4.** QKD link with 4 PM-BPSK classical channels present in the same fiber. Left: no pre-compensation. Right: Pre-compensation with 30 km length of DCF (CD = -132.4 ps/nm/km). Fiber family of Vascade EX2000 fiber was targeted. *Model parameters are listed in Table 1. Legend – see Fig. 2.*

## 6. CONCLUSION

High bit rate 10 Gb/s QKD systems, in contrast to 1 Gb/s systems, are significantly impacted by CD induced ISI in typical fibers. We numerically investigated performance and capacity of high bit rate QKD systems with and without high optical-power classical channels being present by extending the basic model of Eraerds et al. [6] and taking into account CD effects in high bit rate QKD systems. We showed that CD has a significant impact on QKD system's SKR and QBER. We investigated the possibility of CD pre-compensation by DCF or FBG with a fixed accumulated CD and showed that pre-dispersion can essentially achieve SKR and QBER performance of a fiber without CD for a given transmission fiber CD and desired reach**.**

## REFERENCES

[1] Patel, K. A., Dynes, J. F., Choi, I., Sharpe, A. W., Dixon, A. R., Yuan, Z. L., Penty, R. V. and Shields, A. J., "Coexistence of High-Bit-Rate Quantum Key Distribution and Data on Optical Fiber," Phys. Rev. X2, 041010 (2012).
[2] Takesue, H., Nam, S. W., Zhang, Q., Hadfield, R. H., Honjo, T., Tamaki, K. and Yamamoto, Y., "Quantum key distribution over 40 dB channel loss using superconducting single photon detectors," Nature Phot. 1, 343 (2007).
[3] Dixon, A. R., Yuan, Z. L., Dynes, J. F., Sharpe, A. W. and Shields, A. J., "Gigahertz Decoy Quantum Key Distribution with 1 Mbit/s Secure Key Rate," Opt. Express 16, 18790 (2008).
[4] Zhang, Q., Takesue, H., Honjo, T., Wen, K., Hirohata, T., Suyama, M., Takiguchi, Y., Kamada, H., Tokura, Y., Tadanaga, O., Nishida, Y., Asobe, M. and Yamamoto, Y., "Megabits Secure Key Rate Quantum Key Distribution," New J. Phys. 11, 045010 (2009).

[5] Dixon, A. R., Yuan, Z. L., Dynes, J. F., Sharpe, A. W. and Shields, A. J., "Continuous Operation of High Bit Rate Quantum Key Distribution," Appl. Phys. Lett. 96, 161102 (2010).
[6] Eraerds, P., Walenta, N., Legre, M., Gisin, N. and Zbinden, H., "Quantum key distribution and 1 Gbps data encryption over a single fibre," New J. Phys. 12, 063027 (2010).
[7] Patel, K. A., Dynes, J. F., Lucamarini, M., Choi, I., Sharpe, A. W., Yuan, Z. L., Penty, R. V. and Shields, A. J., "Quantum key distribution for 10 Gb/s dense wavelength division multiplexing networks," Appl. Phys. Lett. 104, 051123 (2014).
[8] Choi, I., Zhou, Y. R., Dynes, J. F., Yuan, Z., Klar, A., Sharpe, A., Plews, A., Lucamarini, M., Radig, C., Neubert, J., Griesser, H., Eiselt, M., Chunnilall, C., Lepert, G., Sinclair, A., Elbers, J.-P., Lord, A. and Shields, A., "Field trial of a quantum secured 10 Gb/s DWDM transmission system over a single installed fiber," Opt. Express 22(19), 23121 (2014).
[9] Dynes, J. F., Choi, I., Sharpe, A. W., Dixon, A.R., Yuan, Z. L., Fujiwara, M., Sasaki, M. and Shields, A. J., "Stability of high bit quantum key distribution on installed fiber," Opt. Express 20, 16339 (2012).
[10] E.g., Hangzhou-Shanghai and Nanjing-Suzhou QKD links presented in W. Huang, "QKD in Fiber-Based Optical Networks in China," 5th ETSI/ IQC Workshop on Quantum Safe Security, London (2017).
[11] Downie, J. D., Mlejnek, M., Roudas, I., Wood, W. A., Zakharian, A., Hurley, J., Mishra, S., Yaman, F., Zhang, S., Ip, E. and Huang, Y.-K., "Quasi-Single-Mode Fiber Transmission for Optical Communication," IEEE J. Sel. Topics Quantum Electron. 23, 4400312 (2017).
[12] Korzh, B., Lim, C. C. W., Houlmann, R., Gisin, N., Li, M.-J., Nolan, D., Sanguinetti, B., Thew, R. and Zbinden, H. "Provably secure and practical quantum key distribution over 307 km of optical fibre," Nature Phot. 9, 163 (2015).
[13] Yin, H.-L., Chen, T.-Y., Yu, Z.-W., Liu, H., You, L.-X., Zhou, Y.-H., Chen, Si-Jing, Mao, Y., Huang, M.-Q., Zhang, W.-J., Chen, H., Li, M. J., Nolan, D., Zhou, F., Jiang, X., Wang, Z., Zhang, Q., Wang, X.-B. and Pan, J.-W., "Measurement device independent quantum key distribution over 404 km optical fibre," arXiv:1606.06821 (2016)
[14] Dynes, J. F., Tam, W. W.-S., Plews, A., Fröhlich, B., Sharpe, A. W., Lucamarini, M., Yuan, Z., Radig, C., Straw, A., Edwards, T. and Shields, A. J. "Ultra-high bandwidth quantum secured data transmission," Sci. Reports 6, 35149 (2016).
[15] Peters, N. A., Toliver, P., Chapuran, T. E., Runser, R. J., McNown, S. R., Peterson, C. G., Rosenberg, D., Dallmann, N., Hughes, R. J., McCabe, K. P., Nordholt, J. E. and Tyagi, K. T., "Dense wavelength multiplexing of 1550nm QKD with strong classical channels in reconfigurable networking environments," New J. Phys. 11, 045012 (2009).
[16] Mlejnek, M., Kaliteevskiy, N. A. and Nolan, D. A., "Reducing spontaneous Raman scattering noise in high quantum bit rate QKD systems over optical fiber," arXiv:1712.05891 [quant-ph] (2017).
[17] Arora, O., Garg, A. K. and Punia, S., "Symmetrical Dispersion Compensation For High Speed Optical Links," International Journal of Computer Science Issues 8, 371 (2011)
[18] http://www.laserfocusworld.com/articles/print/volume 33/issue 10/world-news/chirped-bragg-gratings-compensate-for-dispersion.html
[19] http://www.o eland.com/FiberGratingProducts/FiberGrating_dispersion.htm
[20] www.idQuantique.com
[21] Stucki, D., Brunner, N., Gisin, N., Scarani, V. and Zbinden, H., "Fast and Simple One-way Quantum key distribution," Appl. Phys. Lett. 87, 194108 (2005).
[22] Bennett, C. H., and Brassard, G., "Quantum cryptography: public key distribution and coin tossing" in Proceedings of the IEEE International Conference on Computers, Systems, and Signal Processing, Bangalore, India, 175 (1984).
[23] Ma, X., Qi, B., Zhao, Y. and Lo, H.–K., "Practical decoy state for quantum key distribution," Phys. Rev. A 72, 012326 (2005).
[24] https://en.wikipedia.org/wiki/Advanced_Encryption_Standard
[25] Huttner, B., personal communication. ID Quantique (2017)
[26] Proakis, J. G. and Salehi, M., [Digital Communications], McGraw-Hill Higher Education, Boston (2008).
[27] Agrawal, G. P., [Nonlinear Fiber Optics], Academic Press, San Diego (1995).
[28] Stucki, D., Walenta, N., Vannel, F., Thew, R. T., Gisin, N., Zbinden, H., Gray, S., Towery, C. R. and Ten, S., "High rate, long-distance quantum key distribution over 250 km of ultra low loss fibres," New J. Phys. 11, 075003 (2009).